\documentclass[pra,twocolumn,showpacs,superscriptaddress]{revtex4}
\usepackage{graphicx,amsfonts,amssymb,amsmath,hyperref}

\newif\ifhyper
\hypertrue
\ifhyper
\hypersetup{
  citecolor = {green},
  colorlinks = {true}, 
  urlcolor = {blue} 
}
\fi

\newcommand{\Funktion}[2]{#1\kern-0.2em\left(#2\right)}

\newcommand*{\bra}[1]{\mathopen{\langle}#1\mathclose{|}}
\newcommand*{\ket}[1]{\mathopen{|}#1\mathclose{\rangle}}

\begin{document}

\title{Universality of the negativity in the Lipkin-Meshkov-Glick model}

\author{Hannu Wichterich}
\email{hannu@theory.phys.ucl.ac.uk}
\affiliation{
    Department of Physics and Astronomy, %
    University College London, %
    Gower Street, %
    WC1E 6BT London, %
    United Kingdom}
\author{Julien Vidal}
\email{vidal@lptmc.jussieu.fr}
\affiliation{
Laboratoire de Physique Th\'eorique de la Mati\`ere Condens\'ee, CNRS UMR 7600,
Universit\'e Pierre et Marie Curie, 4 Place Jussieu, F-75252 Paris Cedex 05, France
}
\author{Sougato Bose}
\email{sougato@theory.phys.ucl.ac.uk}
\affiliation{
    Department of Physics and Astronomy, %
    University College London, %
    Gower Street, %
    WC1E 6BT London, %
    United Kingdom}

\begin{abstract}
The entanglement between noncomplementary blocks of  a many-body system, where a part of the system forms an ignored environment, is a largely untouched problem without analytic results. We rectify this gap by studying the logarithmic negativity between two macroscopic sets of spins in an arbitrary tripartition of a collection of mutually interacting spins described by the Lipkin-Meshkov-Glick Hamiltonian. This entanglement measure is found to be finite and universal at the critical point for any tripartition whereas it diverges for a bipartition. In this limiting case, we show that it behaves as the entanglement entropy, suggesting a deep relation between the scaling exponents of these two independently defined quantities which may be valid for other systems.
\end{abstract}

\pacs{03.65.Ud; 03.67.-a; 64.70.Tg; 75.10.Jm}

\maketitle 

\section{Introduction}

Considerable attention has recently been devoted to the study of genuinely ``quantum" correlations or entanglement in the ground states of many-body systems among theorists \cite{Amico08} and experimentalists \cite{Ghosh03}, an exciting field that profits from the interplay of quantum information and condensed matter. The measure of entanglement most extensively studied so far is the von Neumann entropy ${\mathcal E}$, which quantifies the entanglement between two complementary parts of a system, the common state of which is  pure. This measure is known to display a universal divergence at criticality \cite{Amico08}. However, the scenario of ``complementary parts" is rather restrictive by demanding that the two parts span the whole of a many-body system. Much more natural is the case where a portion of the system does not belong to either of the parts under consideration; that is, it forms an environment.  Moreover, such a general setting is of relevance for a situation where two parties have access to limited  groups of constituents and attempt to exploit the entanglement between these groups for quantum information tasks. 

In general, the state of two noncomplementary parts $\widetilde{\rho}$ is a statistical mixture and ${\mathcal E}$ is no longer suitable to quantify their entanglement. To this end, {one must invoke} the logarithmic negativity ${\mathcal L}$ \cite{Vidal02}, which is the only adequate measure for this task which is, at the same time, straightforwardly computable. In addition, it has operational meaning (in terms of bounds) in teleportation and distillation. It is defined as ${\mathcal L}=\ln {\rm Tr}((\widetilde \rho^{\text T_1})^\dagger \widetilde \rho^{\text T_1})^{1/2}$, where $T_1$ denotes the partial transposition that amounts to $\bra{m,n}\widetilde \rho^{\text T_1}\ket{k,l}=\bra{k,n}\widetilde \rho \ket{m,l}$ with respect to a complete set of basis states $\ket{k,l}=\ket{k}_1\otimes\ket{l}_2$ on the bipartite system.  Though, for the special case of pure states ${\mathcal L}$ and ${\mathcal E}$ can both be computed from the  so-called Schmidt coefficients \cite{Vidal02}, whether the former is universal and whether its behavior can be related to scaling exponents are open questions. Indeed ${\mathcal L}$ is defined purely from quantum information considerations, and its manifestation of the elegant scaling features from many-body physics will be a true surprise.

Unfortunately, it is notoriously difficult to compute ${\mathcal L}$ even for the simplest of one-dimensional (1D) models for which some numerical results have been obtained \cite{Wichterich09_1,Marcovitch09}.  In this article, we present an analytic study of entanglement, as measured by ${\mathcal L}$, between two macroscopic groups of spins, in a tripartite splitting of a many-body system displaying a quantum phase transition. We mainly focus on the Lipkin-Meshkov-Glick (LMG) model for which we find that ${\mathcal L}$ is always finite  in stark contrast with other macroscopic correlation measures such as mutual information ${\mathcal I}$ which diverges.   Remarkably, ${\mathcal L}$ does not depend on the anisotropy parameter at the transition point and may, in this sense, be considered as universal. However, we also found that this is not a generic feature since the same analysis in the Dicke model \cite{Dicke54} shows a different behavior. Most importantly, we show that {\em in the limiting case of a bipartition, ${\mathcal L}$ diverges at the critical point as $1/6 \ln N$ (where $N$ is the system size), exactly as the entanglement entropy}. This property which is also found in other models leads us to conjecture that it should be valid for all systems.

\section{The Model}

Let us consider a system of $N$ spins $1/2$ which are mutually coupled through  an anisotropic $XY$-type interaction and subjected to a magnetic field of strength $h$ pointing in the $z$ direction. The ground-state entanglement of this model introduced by Lipkin, Meshkov, and Glick (LMG) in 1965 \cite{Lipkin65,Meshkov65,Glick65} to describe nuclei, has been the subject of many recent studies \cite{Vidal04_1,Dusuel04_3,Dusuel05_2,Latorre05_2,Barthel06_2,Orus08_2,Kwok08,Ma08,Quan09}. The LMG Hamiltonian is given by
%
%
\begin{equation}
\label{eq:spinham}
H=-\frac{1}{N}(S_x^2+\gamma\,S_y^2)-h\,S_z,
\end{equation}
%
%
where $S_\alpha =\sum_{k=1}^N \sigma^{(k)}_\alpha/2 \,(\alpha =x,y,z)$,  $\sigma^{(k)}_\alpha$ being the Pauli operators acting on the state space of the $k$th spin. Here, we only consider the case of ferromagnetic interactions and, without loss of generality, we restrict, in a first step,  the anisotropy parameter to $0 \leqslant \gamma < 1$ and the field to $h\geqslant 0$. This system undergoes a second-order quantum phase transition at $h=1$, between a symmetric ($h>1$) and a broken ($h<1$) phase, which is well described by a mean-field approach. The corresponding classical ground state is fully polarized in the field direction ($\langle\sigma_z^i\rangle=1$) for $h>1$, and twofold degenerate with $\langle\sigma_z^i\rangle=h$ for $h<1$ (see Refs.~\cite{Botet83,Dusuel05_2} for details). 

\section{${\mathcal L}$ in a tripartition}

 In order to compute entanglement of the ground state in a tripartite setting, we divide the $N$ spins into three groups 1, 2, and 3 with $N_1$, $N_2$, and $N_3$ spins, respectively, satisfying $N_1+N_2+N_3=N$. Accordingly, we partition the spin operators into $S_\alpha=S_\alpha^{(1)}+S_\alpha^{(2)}+S_\alpha^{(3)}$. To diagonalize the Hamiltonian and hence obtain the ground state in the thermodynamical limit, it is convenient to express the $S_\alpha^{(k)}$'s in terms of bosonic operators using the so-called Holstein-Primakoff representation \cite{Holstein40}
%
%
\begin{eqnarray}
\label{eq:HP1}
S_z^{(k)}&=&N_k/2-a_k^\dagger a_k,\\
\label{eq:HP2}
S_-^{(k)}&=&a_k^\dagger\,\sqrt{N_k}\: (1-a_k^\dagger a_k/N_k)^{1/2}=(S_+^{(k)})^\dagger,
\end{eqnarray}
%
%
where $S_\pm^{(k)}=S_x^{(k)} \pm {\rm i} S_y^{(k)}$. Note that we focus here on maximum spin sectors to which the ground state of Hamiltonian (\ref{eq:spinham}) is known to belong. In addition, as discussed in Ref.~\cite{Dusuel05_2} for the single-mode case, substituting the expressions (\ref{eq:HP1}) and (\ref{eq:HP2}) in (\ref{eq:spinham}) requires a prior rotation of spin operators to bring the $z$ axis along the classical magnetization of the ground state. Using this bosonic representation and expanding $H$ at order $(1/N_k)^0$, one obtains
%
%
\begin{align}
\label{bosonham}
H=\sum_{k,l=1}^3 a^\dagger_k\,A_{k,l}\,a_l+\frac{1}{2}(a^\dagger_k\,B_{k,l}\,a^\dagger_l+\text{H.c.})+ {\rm Cte},
\end{align}
%
%
where $Cte$ denotes constant terms which will be irrelevant in the following and where we introduced the $3 \times 3$ real symmetric matrices
%
%
\begin{equation}
A = r\,\mathbb{I},\quad B= s\,\left(
\begin{tabular}{ccc}
 $\tau_1$ & $\sqrt{\tau_1\tau_2}$ & $\sqrt{\tau_1\tau_3}$  \\
 $\sqrt{\tau_1\tau_2}$ & $\tau_2$ & $\sqrt{\tau_2\tau_3}$ \\
 $\sqrt{\tau_1\tau_3}$ & $\sqrt{\tau_2\tau_3}$ & $\tau_3$
\end{tabular}
\right),
\end{equation}
%
%
with $\tau_{k}=N_{k}/N$ and $\mathbb{I}$ denotes the identity matrix. In the symmetric ($h\geqslant 1$) and broken ($0 \leqslant h< 1$) phases the prefactors read
%
%
\begin{align}
 r&=
\begin{cases}
 \frac{2h-\gamma-1}{2} & h\geqslant 1\,, \\
 \frac{2-\gamma-h^2}{2} & 0\leqslant h< 1\,,
\end{cases}
\\
s&=
\begin{cases}
  \frac{\gamma-1}{2}& h\geqslant 1 \,,\\
  \frac{\gamma-h^2}{2} & 0\leqslant h< 1\,,
\end{cases}
\end{align}
%
%
and we further note that $r>0$ and $r>s$.
Since the Hamiltonian (\ref{bosonham}) is quadratic, it is straightforwardly diagonalized via a Bogoliubov transform. As discussed in \cite{Dusuel04_3} within a single-mode description, the gap for $h\geqslant 1$ is given by $\Delta=\sqrt{(h-1)(h-\gamma)}$, whereas it vanishes as $\exp{(-N)}$ in the broken phase \cite{Botet83}.

\begin{figure}
\includegraphics[width=0.38\textwidth]{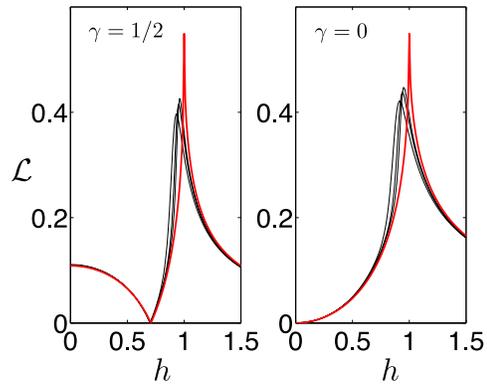}
\caption{(Color online) Logarithmic negativity ${\mathcal L}$ as a function of magnetic field $h$ for an equal tripartition $\tau_1=\tau_2=\tau_3=1/3$ and for two different values of the anisotropy parameter $\gamma$. In the thermodynamic limit , corresponding to the red (gray) lines, at the critical point,  ${\mathcal L}$ is universal (independent of $\gamma$). Black lines correspond to numerical data for $N=90,\,150,\,210$ which match the analytical prediction $N=\infty$ increasingly well. Note also that  ${\mathcal L}$ vanishes for $h=\sqrt{\gamma}$ where the ground state is separable \cite{Dusuel05_2}.
\label{fig.1}}
\end{figure}

Without loss of generality, we compute the entanglement of the ground state $\ket{\psi_0}$ between group 1 and group 3 by computing the logarithmic negativity ${\mathcal L}$ between the corresponding bosonic modes. To capture the entanglement properties of the arising mixed state the only available measure is ${\mathcal L}$ except for the limit of two spins for which concurrence can also be used \cite{Amico08}. In the present context a convenient definition for $\mathcal L$ may be invoked. Indeed, the ground state is a Gaussian state for which an elegant framework is available that has allowed for a multitude of significant analytical results in the past (see, e.g., \cite{Eisert10}). Entanglement between any two of the groups may be inferred from the  covariance matrix $\Gamma$ which collects the second moments $\Gamma_{i,j}=\bra{\psi_0} R_i R_j+ R_j R_i\ket{\psi_0},\,({\rm with} \: i,j=1,2,\ldots,6)$ of canonical coordinates $x_k=(a_k^\dagger+a_k)/\sqrt{2}$ and momenta $p_k={\rm i}\,(a_k^\dagger-a_k)/\sqrt{2}$, which we group together in the vector $R=(x_1,x_2,x_3,p_1,p_2,p_3)$.  In this representation $\Gamma$ adopts an explicit expression in terms of matrices $V_x=(A+B)$ and $V_p=(A-B)$ \cite{Cramer06} which, by virtue of $[V_x,V_p]=0$,  reads $\Gamma=\Gamma_x\oplus\Gamma_p$, where $\Gamma_x=\Gamma_p^{-1}=V_p^{-\frac12}V_x^{\frac12}$. Similarly,  the reduced density operator $\widetilde {\rho}=\text{Tr}_2(\ket{\psi_0}\bra{\psi_0})$ ($\text{Tr}_2$ denotes the partial trace over group $2$) has a representation in terms of a Gaussian state with covariance matrix $\widetilde \Gamma$ obtained from $\Gamma$ upon canceling rows and columns that correspond to mode $2$, namely, $\widetilde \Gamma=\widetilde \Gamma_x\oplus\widetilde \Gamma_p$, where
%
%
\begin{align}\label{covmat}
\widetilde \Gamma_x&=\mathbb{I}+(\alpha^{-1}-1)\left(
\begin{tabular}{cc}
 $\tau_1$ & $\sqrt{\tau_1\tau_3}\,$  \\
 $\sqrt{\tau_1\tau_3}$ & $\tau_3$
\end{tabular}
\right), \\\label{covp}
\widetilde \Gamma_p&=\mathbb{I}+(\alpha-1)\left(
\begin{tabular}{cc}
 $\tau_1$ & $\sqrt{\tau_1\tau_3}$  \\
 $\sqrt{\tau_1\tau_3}$ & $\tau_3$
\end{tabular}
\right),
\end{align}
%
%
and $\alpha=\sqrt{(r+s)/(r-s)}>0$.
Then, upon partial transposition $T_1$, the covariance matrix is subjected to partial time reversal $p_1\rightarrow -p_1$ \cite{Simon00} and is transformed into $\widetilde{\Gamma}^{T_1}$ obtained from $\widetilde{\Gamma}$ by changing the sign in the off-diagonal terms in Eq.~(\ref{covp}). The logarithmic negativity can then be computed in terms of the symplectic (degenerate) eigenvalues  of $\widetilde{\Gamma}^{T_1}$,
%
%
\begin{equation}\label{sympl1}
\lambda_{1,2}=\sqrt{1+g \pm\sqrt{g^2+4 \,\tau_1\tau_3(\alpha+\alpha^{-1}-2)}},
\end{equation}
%
%
where $g=\big[\tau_1+\tau_3-(\tau_1-\tau_3)^2\big]\,\big(\alpha+\alpha^{-1}-2\big)/2\geq 0$ and where the $+$ ($-$) sign corresponds to $\lambda_1$ $(\lambda_2)$. Indeed, noting that $\lambda_1\geqslant 1 \geqslant \lambda_2 \geqslant 0$, the logarithmic negativity reads as \cite{Vidal02}
%
%
\begin{equation}\label{logneg}
 {\mathcal L}=-\sum_{i=1}^2 \ln\text{min}(\lambda_i,1)=-\ln\lambda_2~.
\end{equation}
%
%
Equation (\ref{logneg}) is the central result of this article and we now discuss it in detail. First, let us note that for an arbitrary tripartition of the system ($0<\tau_k<1$, for $k=1,2,3$)  ${\mathcal L}$ is {\em finite} across the whole phase diagram $(h,\gamma)$ including the transition point $h=1$, where one has
%
%
\begin{equation}
\label{eq:L_ani_crit}
{\mathcal L}(\tau_1,\tau_3,\gamma,h=1)=\frac{1}{2}\ln
\frac{\tau_1+\tau_3-(\tau_1+\tau_3)^2}{\tau_1+\tau_3-(\tau_1-\tau_3)^2}.
 \end{equation}
 %
 %
Remarkably, this expression does not depend on the anisotropy parameter $\gamma$ revealing the {\em universal character} of the logarithmic negativity at the critical point of the LMG model. The fact that a measure of quantum correlations between macroscopic groups of particles is finite at a quantum critical point is also entirely nontrivial. For example, the mutual information ${\mathcal I}(1,3)={\mathcal E}(3,1\cup2)+{\mathcal E}(1,2\cup3)-{\mathcal E}(2,1\cup3)$ diverges as, indeed, for the simple case of an equal tripartition, one has ${\mathcal E}(3,1\cup2)={\mathcal E}(1,2\cup3)={\mathcal E}(2,1\cup3)$. Thus mutual information diverges as the entropy, that is, as $\tfrac{1}{6} \ln N$  \cite{Vidal07}, while it was found to be finite in 1D and for short-ranged interactions \cite{Furukawa09,Calabrese09}. As $\mathcal I$ measures all correlations, one may conclude that it is the classical part of correlations that is responsible for the divergence, whereas the quantum part (as measured by ${\mathcal L}$) remains finite. We depict the behavior of ${\mathcal L}$ for such a tripartition and compare it with data from exact diagonalization in Fig.~\ref{fig.1}.  

%
%
%
%

\section{${\mathcal L}$ in a bipartition}

Most importantly, when $\tau_1=\tau$ and $\tau_3=1-\tau$, which corresponds to the limiting bipartite case $\tau_2=0$, ${\mathcal L}$ diverges at the critical point. Such a behavior is in agreement with the fact that, for a bipartite pure state, ${\mathcal L}$ is lower bounded by ${\mathcal E}$ \cite{Vidal02} which is divergent at $h=1$. To analyze this divergence, one expands ${\mathcal L}$ in the vicinity of the critical point by imposing this bipartition condition from the beginning and one obtains~:
%
%
\begin{eqnarray}
{\mathcal L}(\tau,1-\tau, \gamma,h)&=&-\frac{1}{4}\ln |h-1|+\frac{1}{4}\ln (1-\gamma) \\
&&+\frac{1}{2}\ln \tau(1-\tau)+1+O(|h-1|^{1/2}). \nonumber
 \end{eqnarray}
 %
 %

\begin{figure}
\includegraphics[width=0.38\textwidth]{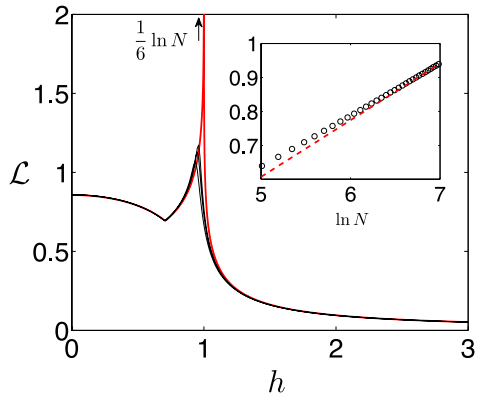}
 \caption{(Color online) Logarithmic negativity ${\mathcal L}$  as a function of the magnetic field $h$ for a bipartition $\tau_1=1/3,\,\tau_3=2/3$, and $\gamma=1/2$ shown as red (gray) line. In the broken phase $h<1$ we plot ${\mathcal L}+\ln 2$ for a reason detailed in the text. Black lines corresponds to exact diagonalization data for $N=180, 270, 360$. Inset: Scaling of ${\mathcal L}$ with system size $N=120,150\ldots,1080$ from exact diagonalization (black circles) approaching a linear dependence on $\ln N$ with slope $1/6$ (dashed line) for large $N$.\label{fig.2}}
\end{figure}

It is interesting to note that this singular behavior is exactly the same (up to constant terms) as the one obtained for other entanglement measures computed in this model \cite{Orus08_2}. This correspondence allows us to straightforwardly extract the finite-size behavior at the critical point by using the same line of reasoning. Indeed, the scaling argument introduced in Refs.~\cite{Dusuel04_3,Dusuel05_2} yields
%
%
\begin{equation}\label{eq:divergence}
{\mathcal L}(\tau,1-\tau, \gamma,1)\sim\frac{1}{6}\ln N+\frac{1}{6}\ln (1-\gamma) \\
+\frac{1}{2}\ln \tau(1-\tau).
\end{equation}
 %
 %
In order to check this behavior, we perform exact diagonalization for increasing system sizes at $h=1$. As can be seen in  Fig.~\ref{fig.2}, numerical data perfectly match the analytical predictions of the thermodynamical limit. Note that, in the broken phase, there is an offset of $\ln 2$ which is due to the fact that the ground state is twofold degenerate. As can be easily understood in the  limit $h,\gamma \rightarrow 0$ for which the finite-size numerical ground state is given by a cat-state, this offset is only present for a bipartition but does not occur for a tripartition, as can be seen in Fig.~\ref{fig.1}. The expression (\ref{eq:divergence}) allows us to add one more equivalence of critical scaling laws in the LMG model since, at the critical point, we have now
%
%
\begin{equation}
{\mathcal G} \sim {\mathcal S} \sim {\mathcal E} \sim {\mathcal L},
 \end{equation}
 %
 %
where ${\mathcal G}$ is the geometric entanglement, ${\mathcal S}$ the single-copy entanglement, ${\mathcal E}$ the entanglement entropy, and ${\mathcal L}$ the logarithmic negativity \cite{Orus08_2}. Of course, it would be very valuable to establish the same kind of equivalence in 1D spin systems for which one already knows that \cite{Orus08_2}
%
%
\begin{equation}
\frac{1}{N}{\mathcal G} \sim \frac{1}{2}{{\mathcal S}} \sim \frac{1}{4} {\mathcal E}.
 \end{equation}
 %
 %
 
%
%
%
%

Although a rigorous proof is still missing, our result together with recent numerical studies\cite{Wichterich09_1,Marcovitch09} leads us to conjecture that in 1D critical systems and for a bipartition, one has ${\mathcal E} \sim {\mathcal L}$. If confirmed, this result may even be valid in any dimensions.

\section{The isotropic case}

 Finally, let us discuss the case $\gamma=1$ which is trivially solved since \mbox{$H(\gamma=1)$} commutes with $\mathbf{S}^2$ and $S_z$ so that the eigenstates are the (permutation-symmetric) Dicke states $|S,M\rangle$. For $h>1$, the ground state is fully polarized in the $z$ direction ($S=N/2$ and  $M=N/2$) and, consequently, ${\mathcal L}=0$ for any tripartition. For $h<1$, the nondegenerate ground state is still in the maximum spin sector $S=N/2$ but $M$ decreases with $h$ \cite{Dusuel05_2}. The isotropic case is thus in a different universality class as compared to $\gamma \neq 1$ and it is interesting to compute ${\mathcal L}$ in the limit $h\rightarrow 1^-$. There,  the ground state is given by $|S,M\rangle=\ket{N/2,N/2-1}$ whose logarithmic negativity between groups $1$ and $3$ reads
%
%
\begin{equation}
{\mathcal L}(\tau_1,\tau_3,1,h\rightarrow 1^-) =
\ln\Big(1- \tau_2+\sqrt{\tau_2^2+4\tau_1\tau_3}\Big).
\end{equation}
%
%
%
This expression strongly differs from Eq.~(\ref{eq:L_ani_crit}) (different universality class) while agreeing with the universal character of ${\cal L}$ in the LMG model at criticality.

\section{Discussion}

The present study reveals two main properties of the logarithmic negativity at a critical point~: ({\it i}) for a tripartition $\mathcal{L}$ is universal and finite~; ({\it ii}) for a bipartition $\mathcal{L}$ is universal and diverges as $\mathcal{E}$. To check the generality of these results, we computed $\mathcal{L}$ in the Dicke model \cite{Dicke54} for which the ground-state entropy has been already computed \cite{Lambert04,Vidal07}. This model describes a set of $N$ spins $1/2$ interacting with a single-mode bosonic field via the Hamiltonian . Thus, if one divides the spins in two parts, one can consider two different negativities (spin-spin or field-spin). We computed both quantities and we found that, contrary to the LMG model, for a tripartition, $\mathcal{L}$ is not universal (but still finite) at the critical point. However, in the bipartition limit, we found that $\mathcal{L}$ behaves also as $\mathcal{E}$ at the transition point.

These complementary studies of the Dicke and LMG models together with 1D spin chain analysis \cite{Wichterich09_1,Marcovitch09} lead us to conjecture that {\it for a tripartition $\mathcal{L}$ is finite but not universal}, even at the critical point. Furthermore, in the bipartition limiting case, {\it $\mathcal{L}$ and $\mathcal{E}$ behave similarly at the transition point}.

A very challenging question would be to check the veracity of this conjecture in other spin systems, in particular in 1D where conformal field theory approaches may allow for exact results.\\

\begin{acknowledgments}
 We thank M. Cramer, S. Dusuel, and A. Serafini for very helpful comments. HW is supported by the EPSRC, United Kingdom. SB acknowledges the EPSRC, United Kingdom; the QIPIRC; the Royal Society; and the Wolfson Foundation.
\end{acknowledgments}


\end{document}